\documentclass[a4paper]{article}
\pdfoutput=1
	\usepackage{fullpage}
	\usepackage{amssymb}
	\usepackage{amsmath}
	\usepackage{graphicx}
	\usepackage{color}
	\usepackage{subfigure}
	\usepackage{array}
	\usepackage{tabularx}
	\usepackage{multirow}
	
	\newlength{\figurewidth}
	\newlength{\figurewidthA}
\newcommand{\dd}{\textrm{d}}

\newcommand{\refeq}   [1] {(\ref{#1})}
\newcommand{\reffig}  [1] {Fig.~\ref{#1}}
\newcommand{\refTab}  [1] {Table~\ref{#1}}
\newcommand{\refSect} [1] {Sect.~\ref{#1}}
\newcommand{\refappe} [1] {appendix~\ref{#1}}
\newcommand{\refAppe} [1] {Appendix~\ref{#1}}

\begin{document}


\begin{flushleft}
{\Large
\textbf{Measuring the dimension of partially embedded networks}
}
\\
D\'aniel Kondor$^{1,\ast}$,
P\'eter~M\'atray$^{2}$,
Istv\'an Csabai$^{1}$,
G\'abor Vattay$^{1}$ 
\\
\bf{1} Department of Physics of Complex Systems, %
E\"otv\"os Lor\'and University, H--1117 Budapest,  P\'azm\'any P. s. 1/A, Hungary
\\
\bf{2} Department of Information Systems, E\"otv\"os Lor\'and University, Hungary\\
H-1117 Budapest, P\'azm\'any P\'eter S\'et\'any 1/C
\\
$\ast$ E-mail: kdani88@elte.hu
\end{flushleft}


%
%
%
%

\section*{Abstract}
Scaling phenomena have been intensively studied during the past decade in the context of complex networks. As part of these works, recently novel 
methods have appeared to measure the dimension of abstract and spatially embedded networks. In this paper we propose a new dimension measurement 
method for networks, which does not require global knowledge on the embedding of the nodes, instead it exploits link-wise information (link 
lengths, link delays or other physical quantities). Our method can be regarded as a generalization of the spectral dimension, that grasps the 
network's large-scale structure through local observations made by a random walker while traversing the links. We apply the presented method to 
synthetic and real-world networks, including road maps, the Internet infrastructure and the Gowalla geosocial network. We analyze the theoretically and 
empirically designated case when the length distribution of the links has the form $P(\rho) \sim 1/\rho$. We show that while previous dimension 
concepts are not applicable in this case, the new dimension measure still exhibits scaling with two distinct scaling regimes. Our observations 
suggest that the link length distribution is not sufficient in itself to entirely control the dimensionality of complex networks, and we show that 
the proposed measure provides information that complements other known measures.

%
%
%
%

\section{Introduction}

\label{sec_intro} The dimensionality of a physical system is an essential parameter reflecting its spatial
scaling properties. The dimension
influences the behavior near a critical point, affecting the scaling of various static and dynamic physical quantities~\cite{rgcrit,dyncrit}.
Magnetic systems can be considered as a classic example, but in the last decades the concepts and methodology of critical 
behavior have been successfully applied to macroscopic and real-world inspired systems too \cite{forest,traffic}. Recently, the connection between 
criticality and dimensionality has also been extensively studied in the context of complex networks~\cite{critnet,qg}.

For general abstract networks however, it is not straightforward to obtain a proper definition of dimensionality.
During the last several years, there appeared a number of methods that were successful in identifying scaling laws in complex networks,
giving suitable generalizations of existing concepts of dimensionality~\cite{spectral, hwang, spectral2, fractality, graph_boxcount, fractal_sw}.
Recently, these methods were also adapted to include global spatial information present in spatially embedded networks~\cite{daqing}.

In this paper we investigate the 
feasibility of a dimensionality measurement for complex networks that relies merely on \emph{local} information. To do so, we give a
generalization of the spectral dimension~\cite{spectral} to record what a random walker ``perceives'' while traversing the network.
While a random walk is not a realistic model for every possible process taking place on a network, it is suitable to gain
some information about the structure of the network. Both the random walk process and the concept of spectral dimension have
been successfully applied to networks previously~\cite{spectral2,diffnet}. Now, we refine the concept of spectral dimension
to include the information present in distances or delays associated with the links, gaining a more complete measure of
network dimensionality.
Our new approach is readily applicable to spatially embedded networks, and additionally it allows the treatment of partially embedded
networks, where the embedding information is only present as link-wise properties (e.g. distances or delays). {These networks can
be considered a special kind of weighted networks, where the link weights are related to the time needed to traverse the link.
Note that we use the weights as generalized distances, which is fundamentally different from the treatment of networks with
arbitrary weighted links~\cite{hwang}.} The motivation for considering such networks is twofold.

First, for many real-world networks the spatial embedding is only partially feasible in practice. For instance, let us suppose a traversal or 
transport process taking place on a complex network lacking spatial information. As a result, we can generally obtain some sort of local, 
link-wise information (e.g.~``delivery times'' along the links), but without gaining any global knowledge on the physical layout of the network. 
In such situations, we can assign the measured link-wise property to each link to obtain a \emph{partial embedding} of the network. A 
straightforward illustration is the Internet, where delays are relatively easy to measure, but reliably determining the geographic 
position of the nodes is not feasible on a large scale \cite{spotter}. For similar partially embedded networks, a random walk processes can 
effectively utilize local knowledge to characterize the large-scale structure of the system.

Another motivation is for ``fully'' embedded networks. Even if there exists a natural (2 or 3D) embedding space for the network in question, it 
may still be relevant to study the network's scaling behavior via local, link-wise properties. As an example, imagine a typical road network, 
where different types of roads have different speed limits. In such a network, travel times are not in a simple relationship with the metric 
distance of the embedding space (the length of road segments between two intersections). It is meaningful to investigate scaling in the light of 
the ``overlay'' property (travel time), instead of the metric distance.

Our method can be considered as an extension of the classical diffusion problem. On an abstract network
the diffusion process is usually considered as the function of the discrete time steps taken (i.e. the number of
hops). In the case of an embedded network however, there is a natural time-scale and the delays suffered while
traversing the links are directly related to the structural properties of the network. If the link delays
have a broad distribution, the dynamics of the resulting process will significantly differ from the standard
diffusion where each step takes the same amount of time.

The rest of the paper is organized as follows. In \refSect{sec_relwork} we give a general overview on network dimension measurements 
and in \refSect{sec_method} we review those that utilize a random walk process. We introduce our method for partially embedded networks 
in \refSect{sec_tau}. In \refSect{sec_results}, we present simulation results for synthetic and real-world networks, and compare the
findings for the presented methods. Finally, we conclude the paper in \refSect{sec_discussion}.

%
%
%
%

\section{Related Work}
\label{sec_relwork}

In the last several years, a number of methods have appeared in the literature that were successful in identifying scale-invariant properties in 
small-world complex networks. Probably, the earliest such concept is the \emph{spectral dimension}~\cite{spectral} which originates from random 
walks on the network (see \refSect{sec_method}), and was applied to both theoretical models of networks~\cite{hwang} and empirical datasets~\cite
{spectral2}. Additionally, the application of the box-counting dimension~\cite{chaosbook} to networks was also proposed~\cite{graph_boxcount}, and 
further generalized to reveal fractal properties of complex networks~\cite{fractality,fractal_sw}. The scaling exponents arising from these 
methods are usually interpreted as a special type of network dimension.

Beyond these methods, which handle networks as abstract graphs, there have been an increasing interest in including the spatial 
properties of networks, too (see e.g.~\cite{barthelemy,yook,lambiotte, nowell, gowalla, matray,distance}). Particularly, in the context of 
dimension measurements, the presence of spatial information enables the application of well-known approaches \cite{correldim, chaosbook} 
to determine the fractal dimension of the point set of network nodes \cite{yook}. A shortcoming of these approaches may be, that while they 
take into account the geometric layout of the network, they entirely neglect its connectivity information. Recently, in \cite{daqing} Daqing
\emph{et al.}~have proposed more suitable methods to overcome this limitation. The authors combine \emph{both} metric and topological knowledge to 
yield more comprehensive measures of dimensionality.

%
%
%
%

\subsection{Random walks and dimensionality of graphs} 
\label{sec_method}

A principal method to define the dimensionality of an abstract graph is performed by examining the properties of a random walk process. We define 
a graph $G$ with a set of nodes $V$ and a set of edges $E$ between the nodes. An edge is an unordered pair $(i,j)$ where $i$ and $j$ are distinct 
nodes from $V$ (the edges are undirected, and we do not allow self-edges). The graph can be represented by its adjacency matrix $\mathbf{A}$, 
where $A_{ij}=A_{ji}=1$ if there is an edge between $i$ and $j$, and otherwise $0$. Let $\mathbf{D}$ denote the degree matrix of $G$, the diagonal 
matrix with entries $D_{ii} = \sum_{j} A_{ij}$.

%
%
%
%

The spectral dimension can be estimated from the scaling behavior of the probability that a random walker returns to the origin of the
walk~\cite{spectral,havlin_diffusion}. We can define the transition probability from node $i$ to $j$ in $t$ steps: 
\begin{equation}
p_{ji}(t) = (\mathbf{T}^t)_{ji} \textrm{,}
\end{equation}
%
where $\mathbf{T} = \mathbf{A}\mathbf{D^{-1}}$ is the transition matrix of the process.



We will be interested in the $P_0(t) = \langle p_{ii}(t)\rangle_i$ \emph{return probability} of the walk, that is the average probability that the
random walk returns to the origin after $t$ steps (the average goes over all possible $i$ starting nodes).
If the return probability exhibits a 
\begin{equation}
\label{spectraldim}
P_0 \sim t^{-\alpha}
\end{equation}
scaling for a sufficiently large range, we define the spectral dimension of $G$ as $d_s = 2\alpha$. If $G$ is a $d$-dimensional regular lattice,
then $d_s = d$ \cite{spectral,havlin_diffusion}.

We note that the spectral dimension concept is closely related to the spectral density function of the transition matrix.
In the continuous limit, $P_0 (t)$ arises as the Laplace-transform of the spectral density~\cite{spectral,brspectrum}.
Indeed, the spectra of various matrices associated with a network have been intensively studied, also in the context of
diffusion~\cite{brspectrum,derenyi,spectra2,scalefreespectra} and the spectral dimension was found to be a valuable tool
to describe topological properties of real-world networks~\cite{spectral2}. A more exotic case is that of quantum gravity, where the
spectral dimension of the networks defined by the possible triangulations of space-time can be interpreted as the perceived
dimension of the Universe~\cite{qg}.

%
%
%
%


A possible generalization of $d_s$ to spatially embedded networks is given in~\cite{daqing}. A graph $G$ is said to be embedded into a metric 
space $(X,\rho)$ if each node in $V$ corresponds to a point in $X$, and $\rho$ is a metric on $X$, i.e. for each pair of nodes $i$ and $j$ we have 
a distance $\rho(i,j)$. In a general setting $\rho$ is the two or three dimensional Euclidean distance, but for many large-scale real world 
networks it is the spherical distance that plays the role of $\rho$.

In case of an embedded graph, the random walk process can be interpreted as a diffusion in the embedding space $X$. Consequently, we can measure the
exponent of the diffusion on the embedded graph via the scaling relation
\begin{equation}
\label{diff}
r \sim t^\beta \textrm{,}
\end{equation}
where $r(t)$ is the root mean square (r.m.s.) displacement of the random walker at time $t$:
\begin{equation}
r^2(t) = \frac{1}{n} \sum_{ij}{\rho^2 ( i,j ) p_{ji}(t)} \textrm{.}
\end{equation}
The diffusion exponent is $\beta = 0.5$ for regular lattices in any dimension, while for real world systems it often exhibits anomalous behavior
with $\beta \neq 0.5$ \cite{havlin_diffusion}.

The spectral dimension concept employed by Daqing \emph{et~al.}~\cite{daqing} can be extracted from the scaling relation
\begin{equation}
\label{daqing}
P_0 (t) \sim r^{-\gamma} \textrm{,}
\end{equation}
where $r = r(t)$. Here, the exponent gives an alternative measure for the dimension of the network: $d_{\rho} = \gamma$. In the case where the
three scaling laws (Eqs.~\refeq{spectraldim},~\refeq{diff} and~\refeq{daqing}) are all valid in the same range, the three exponents are
related: $\gamma = \alpha / \beta$. For regular $d$ dimensional lattices this relationship is satisfied as $d_{\rho} = d_s = d$ and $\beta = 1/2$.
Nevertheless, for more complex networks, the scaling regimes may not coincide, or some of the scaling relationships might not hold at all.

%
%

\subsection{The role of the link length distribution}
\label{sec_lld}

A remarkable result of \cite{daqing} is that the authors demonstrate, that the $P(\rho)$ distribution of link lengths has a central role in 
controlling the dimensionality of a spatial network. Inspired by that result, we also pay special attention to the distribution of link
lengths of the networks considered. We emphasize, that the distribution of link lengths can be defined in two different ways.
Throughout this paper, by $P(\rho)$ we refer to the \emph{observed} distribution, i.e.~the probability that a link in the
network has length $\rho$. Instead of $P(\rho)$, many authors also use the $P_{(c)} (\rho)$ \emph{conditional} version:
the probability that two nodes are connected given that they are separated by distance $\rho$.
The two definitions are not independent: $P_{(c)} (\rho) = P(\rho) / C(\rho)$, where $C(\rho)$
is the distribution of distances among any two points regardless of whether there is a link between them or not. For a set of uniformly distributed points
in $d$ dimensions, we have $C(\rho) \sim \rho^{d-1}$ and thus $P_{(c)} (\rho) \sim \rho^{1-d} P(\rho)$. On the other hand, the nodes in a real-world
spatially embedded network are in many cases inhomogeneously distributed, which complicates the relation between $P(\rho)$ and $P_{(c)} (\rho)$. To this
end, we only use the observed distribution $P(\rho)$ as it can be estimated directly from the data.

Based on empirical observations, $P(\rho)$ is often assumed to follow a power law decay:
$P(\rho) \sim \rho^{-a}$~\cite{lambiotte, nowell, matray, distance}.
The special case of $a = 1$ has been of interest from both theoretical and empirical perspective. In his seminal paper~\cite{kleinberg},
Kleinberg showed that for an important family of small-world networks the $a$ exponent is controlling the navigability of the network: 
at $a=1$ efficient routing is achievable based merely on local information, while if $a \neq 1$, this is not feasible.
(Note, that Kleinberg used the $P_{(c)} (\rho)$ conditional version of the link length distribution and thus stated his results for
$P_{(c)} (\rho) \sim \rho^{-2}$ in two dimensions.)
Kleinberg's results are in agreement with the empirical finding that in many real-world networks, $a \approx 1$. Early evidence for the Internet 
was found by Yook \emph{et al.}~\cite{yook}, which has also been recently supplemented by more precise measurements by the authors of the present 
paper~\cite{matray}. The same phenomenon was also observed in social networks~\cite{nowell}, mobile communication networks~\cite{lambiotte}, and 
e-mail networks~\cite{distance}. Very recently, in~\cite{humodell} Hu \emph{et~al.}~proposed a statistical model which reproduces this peculiar 
scaling phenomenon. The authors found that the $P(\rho) \sim 1 / \rho$ behavior arises naturally from an entropy-maximization constraint. The 
Internet infrastructure and the geosocial network used in our analysis can also be well characterized by $a \approx 1$ (see \reffig{fig_lld}).

%
%
%
%

\section{Partially embedded networks}
\label{sec_tau}

\begin{figure}[tbp]
\begin{center}
\includegraphics[width=\figurewidth, angle=0]{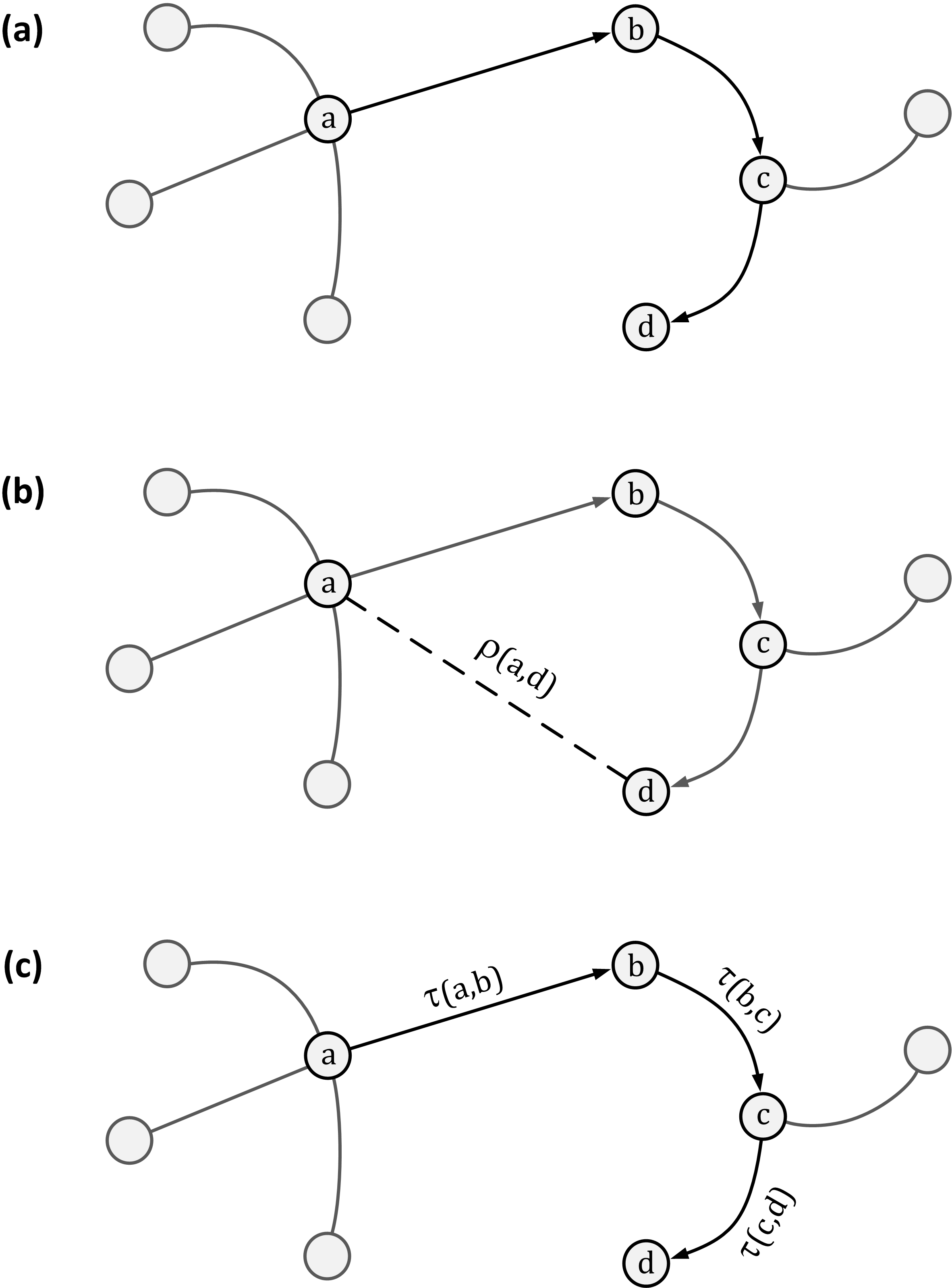}
\caption{Different methods to measure scaling via the random walk process. (a) shows a specific walk $s = (a, b, c, d)$ on an abstract network in $t=3$. Here, the $d_s$ spectral dimension can be calculated via the Monte Carlo simulation of Eq.~\refeq{spectraldim}. (b) If the network is embedded into a metric space, we can obtain $\rho(a,d)$ (and generally $\rho(\cdot,\cdot)$ for any node pair), hence it is feasible to determine $d_{\rho}$ via Eq.~\refeq{daqing}. (c) In case the network is only partially embedded, $\rho(a,d)$ is not available generally, and thus it is not possible to obtain $d_{\rho}$. However, if a $\tau$ length (or delay) function is present for the links, we are able to cumulate the $\tau$ values along the walk to obtain $d_{\tau}$ via Eq.~\refeq{taudim}. In this particular case $l(s) = \tau(a,b) + \tau(b,c) + \tau(c,d)$.}
\label{fig_randomwalk}
\end{center}
\end{figure}

In the following, we propose a new method to measure the dimensionality of spatially embedded networks, which can also be generalized to the special
type of networks which we refer to as \emph{partially embedded}.

This class of networks lies in-between abstract networks (with topological information only) and spatial networks (with both topological and 
embedding information). These networks emerge either when the $\rho(i,j)$ distances are not available generally, but only for connected node pairs
(i.e.~when $(i,j) \in E$), or when we wish to study the network as per a link-wise overlay property possibly different from $\rho$ (e.g. 
link-wise delays). This relaxation yields networks that are spatially constrained to some extent, but the embedding information is incomplete, as 
the nodes do not have coordinates. Due to their transient character, we refer to these networks as \emph{partially embedded networks} \footnote
{Note that the term ``partially embedded'' is also used in the mathematics literature for graphs that have a subgraph embedded in the plane.}. To 
distinguish spatially embedded networks from partially embedded networks, we refer to the link length function in a partially embedded network
as $\tau(i,j)$, which needs to be defined only for the edges of the network~\footnote{As both $\rho$ and $\tau$ can denote the length of a 
link, we use $\tau$ to emphasize when the distance needs to be defined only for the edges of the network. Also, for a spatially embedded network 
we can define $\tau$ as an arbitrary function of $\rho$.}.

To estimate the dimensionality of such networks, we consider random walks similarly to \refSect{sec_method}, and exploit the inherent information 
in the $\tau$ measure. Let $s = (s_0, s_1, \ldots, s_m)$ denote a specific walk of length $|s| = m$ from node $s_0$ to node $s_m$. We can measure 
the cumulated length of a walk $s$ consisting of $m$ steps as
\begin{equation}
	\label{traveltime}
	l(s) = \sum_{i=0}^{m-1} \tau(s_i, s_{i+1}) \textrm{.}
\end{equation}
Considering only walks which return to the origin, we denote the distribution of the lengths of such walks by $P(l)$.
We are then interested in the scaling relation:
\begin{equation}
	\label{taudim} P(l) \sim l^{-\delta} \textrm{,}
\end{equation}

where the proposed dimension measure can be estimated as $d_\tau = 2\delta$ (if $P(l)$ exhibits scaling in a suitably
large range). Note, that this definition resembles that of the spectral dimension. Indeed, Eq~\refeq{taudim} can be
interpreted as a generalization of Eq~\refeq{spectraldim}. Assuming that the random walker travels with unit velocity,
it is natural to interpret the $\tau(u,v)$ distance as the ``elapsed time'' as perceived by the random walker while
traversing the link between $u$ and $v$. Thus, by $d_\tau$ we achieve a dimension concept that grasps the network's
large-scale spatial structure through \emph{local} characteristics observed by a random walker. Trivially, for a
regular $d$-dimensional lattice $l \sim t$, which gives $d_{\tau} = d_{s} = d$. On a network, where the
distribution of $\tau$ link lengths has a finite second moment, the $l \sim t$ relation will be a good approximation
for large $t$ because of the central limit theorem. If the link length distribution is broad (no finite second moment,
e.g. a power law with an exponent $a \leq 3$), this will not be true; even for large $t$, there will be a significant
difference based on whether we examine the diffusion process as a function of steps taken or elapsed time. Our
goal with the new dimension measure is to take these effects into account. The connection between the return
probability and $P(l)$ is further discussed in Appendices~\ref{ap_p0} and~\ref{ap_dtauprop}. In \reffig{fig_randomwalk}
we illustrate the different approaches applied to measure a network's dimension: $d_s$ for abstract networks,
$d_{\rho}$ for spatially embedded networks, and $d_{\tau}$ for partially embedded networks.

%
%
\section{Results}
\label{sec_results}

To study the scaling relations described in Sects.~\ref{sec_method} and \ref{sec_tau}, we considered simulations of random walks on synthetic and 
real-world networks. In accordance with \refSect{sec_lld} we specifically focused on the distribution of link lengths in the networks. All 
networks considered here have full embedding information: the distance $\rho(i,j)$ can be calculated for any $(i,j)$ pair of nodes, and also $\rho$
qualifies as a metric. In accordance with that, when calculating $d_{\tau}$ we simply set $\tau(i,j) = \rho(i,j)$ for node pairs $(i,j)$ for 
which there is a link. This allows us to easily compare the two dimension measures $d_{\rho}$ and $d_{\tau}$.

\subsection{Synthetic spatial networks}
\label{sec_kl}

\begin{figure}[tbp]
  \centering
  \subfigure[Estimating $d_s$]{\label{fig_klspd} %
  \includegraphics[width=\figurewidthA, angle=0]{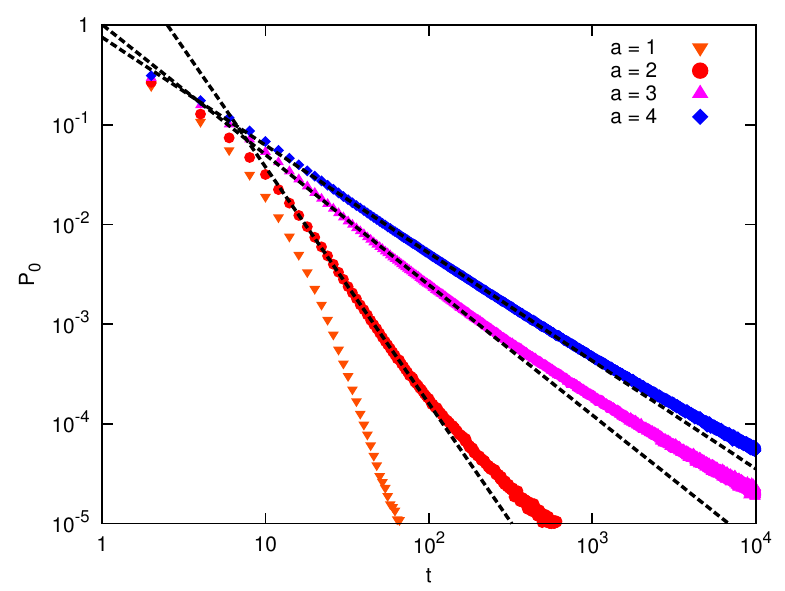}
  }
\quad
  \subfigure[Estimating $d_{\rho}$]{\label{fig_kldqd} %
  \includegraphics[width=\figurewidthA, angle=0]{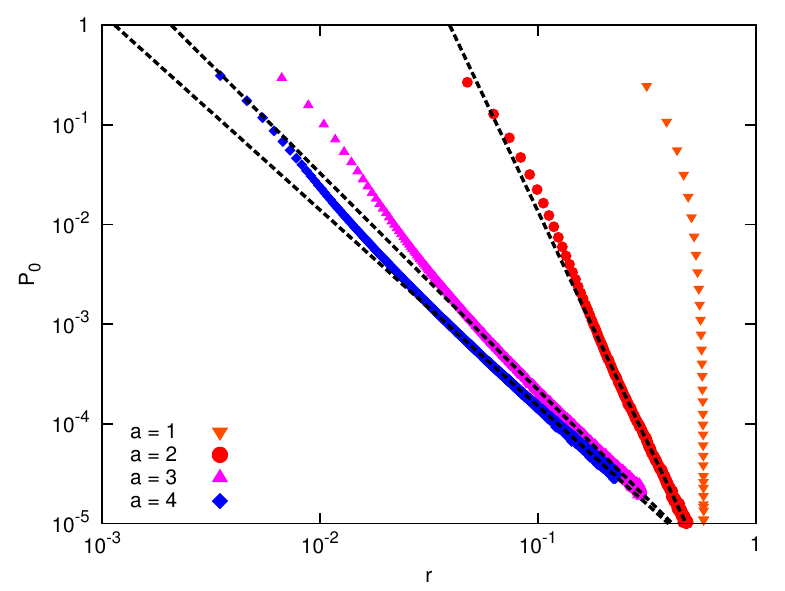}
  }
  \caption{Two different dimension measurement methods for synthetic networks with $a=1, 2, 3$ and $4$.
  (a) The $d_{s}$ spectral dimension (Eq.~\ref{spectraldim})
  		gives scaling in a substantially large range, with monotonically decreasing exponents.
		The resulting dimensions are $d_{s} = 4.75 \pm 0.02, 2.61 \pm 0.01$ and $2.163 \pm 0.003$ for the parameters $a = 2, 3$ and
		$4$ respectively (the lines are fitted in the ranges $t \in  [ 12,81 ] $, $t \in  [ 7,148 ] $ and $t \in  [ 7,403 ] $).
		In the $a = 1$ case we cannot identify scaling, the return probability
		decreases faster than a power-law, which can be interpreted as an infinite dimension.
	(b) The $d_{\rho}$ dimension (Eq.~\ref{daqing}). Here we have increasing exponents again as
		$a$ decreases, while in the $a = 1$ case there is apparently no scaling regime, which again
		indicates infinite dimension. The estimated dimensions are $d_{\rho} = 4.61 \pm 0.01, 2.174 \pm 0.001$ and
		$1.967 \pm 0.001$ for $a = 2, 3$ and $4$. The lines were fitted for the ranges $r > 0.183$, $r > 0.03$ and
		$r > 0.015$ for $a = 2, 3$ and $4$.
	}
  \label{fig_dims}
\end{figure}

We generated spatially embedded random networks with a given link length distribution, following the 
method of~\cite{kosmidis}. First, we construct a $d = 2$ dimensional regular lattice, and then for each node $u$ we draw a random degree $d_u$ 
from a Poisson distribution with mean $k$ (note, that this process leads to a network with an average degree $\approx 2k$). Next, for each $d_u$
links of $u$ we generate a random length $\rho_{1}$ from $P(\rho)$, and link to a random node $v$ such that $\rho(u,v) \approx \rho_{1}$.
We scale the lattice such that all coordinates fall between $0$ and $1$. Considering a lattice with linear size $m$, we have $N = m^d$ nodes.
Consequently, the minimum distance between any two nodes is $\rho_{\textrm{min}} = 1/(m-1)$, while the maximum distance is
$\rho_{\textrm{max}} = \sqrt{d}$. In the networks generated, we used power-law distributions $P(\rho) = C \rho^{-a}$, where $C$ is a normalizing
constant such that $\int_{\rho_{\textrm{min}}}^{\rho_{\textrm{max}}} P(\rho) \dd \rho = 1$. Note that since $P(\rho)$ has a finite support, we can
use any exponent $a$ (the distribution can be normalized even for $a \leq 1$). For our simulations we used $a = 1,2,3$ and $4$.

\begin{figure}[tbp]
  \centering
  \subfigure[Estimating $d_{\tau}$ ($a = 2,3,4$)]{ \label{fig_kltau24}
  \includegraphics[width=\figurewidthA, angle=0]{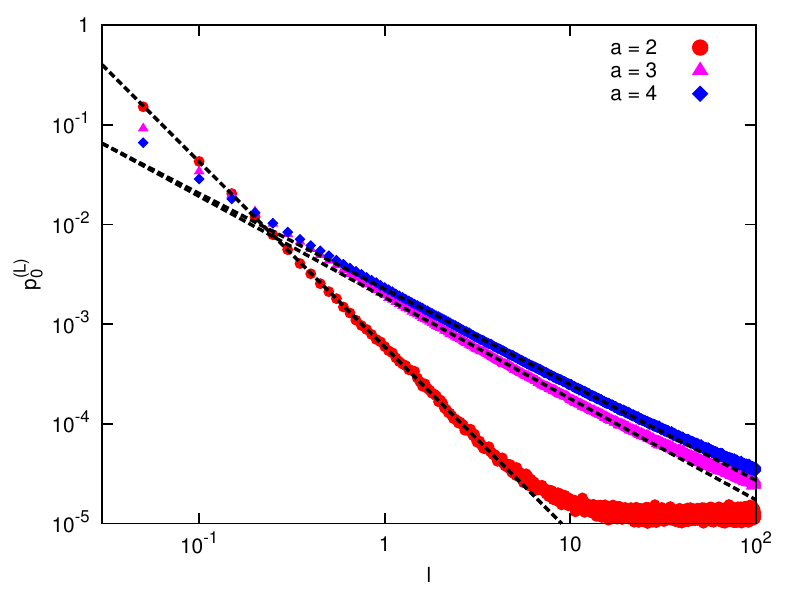}
  }
\quad
  \subfigure[Estimating $d_{\tau}$ ($a = 1$)]{\label{fig_kltau1} %
  \includegraphics[width=\figurewidthA, angle=0]{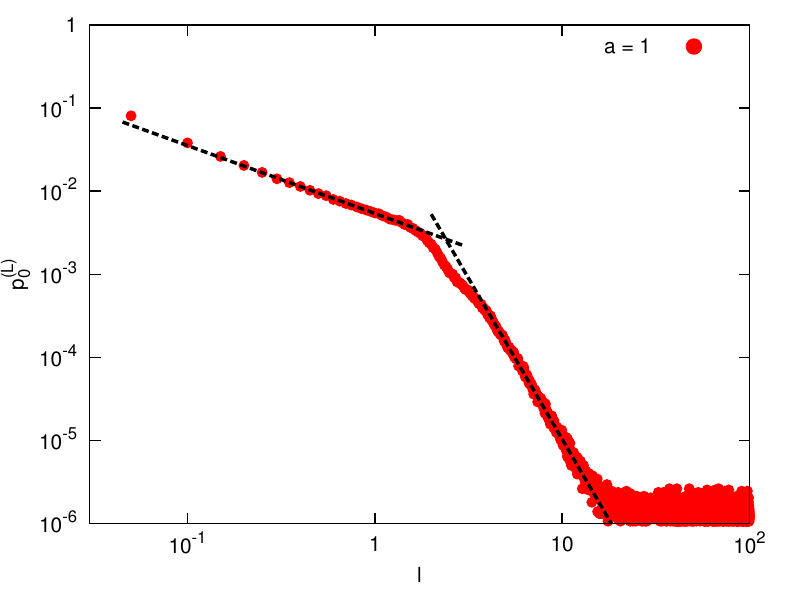}
  }
  \caption{The $d_{\tau}$ dimension measure for synthetic networks with different link length distributions.
  (a) For $a = 2, 3$ and $4$ the estimated dimensions are $3.71 \pm 0.01$, $2.029 \pm 0.004$ and $1.918 \pm 0.003$ respectively
  (the lines were fitted in the ranges $l \in  [ 0.1,3 ] ,  [ 0.4,20 ] $ and $ [ 0.4,20 ] $ for $a = 2, 3$ and $4$). %
  (b) For $a = 1$ there emerge two distinct scaling regimes. We have $d_{1} = 1.63 \pm 0.01$ for the $l \lesssim \rho_{\textrm{max}}$
  ($l \in  [ 0.01,1.35 ] $) range and $d_{2} = 7.75 \pm 0.02$ for the $l \gtrsim \rho_{\textrm{max}}$ ($l \in  [ 3.15,8.2 ] $) range.}
  \label{fig_dtau}
\end{figure}

In \reffig{fig_klspd} we plot the return probability as a function of the steps taken for a specific realization of the spatially embedded random 
network. Note that in this case we only used the connectivity information -- the spatial embedding and $P(\rho)$ only affect the results through 
the random generation of the networks. For $a \geq 2$ we get increasing ranges of scaling regimes, with decreasing estimates of $d_s$. This is in 
accordance with the fact that with increasing values of $a$, there will be fewer long-range links and the network will be dominated by short-range links, 
which augment the underlying two dimensional structure. In the $a=1$ case, we have no scaling -- the return probability can be well approximated 
with the stretched exponential function $P_0(t) \sim \exp{\left( c t^{\epsilon} \right) }$, which is also found in uncorrelated random networks
(Erd\H os -- R\'enyi networks), and in glassy systems~\cite{brspectrum}. We can interpret this result as the network having an infinite spectral
dimension.

A similar phenomenon can be observed for the $d_{\rho}$ dimension in \reffig{fig_kldqd}. Networks with larger values of the $a$ exponent give good 
scaling relations and decreasing dimension estimates, while for $a = 1$ the broad link length distribution yields no scaling, which can again be
regarded as infinite dimension.

We display the results for the new dimension measure $d_{\tau}$ in \reffig{fig_dtau}. In the $a \geq 2$ cases (\reffig{fig_kltau24}) we get 
apparent scaling behavior in all cases, with decreasing dimensions. However, in the $a = 1$ case, we get an anomalous behavior. As seen on
\reffig{fig_kltau1} a separation of length scales emerges at approximately the size of the system ($\rho_{\textrm{max}} = \sqrt{2}$ in the case of our 
synthetic random graphs). We can identify two scaling exponents: $\delta_{1}$ for $l \lesssim \rho_{\textrm{max}}$ and $\delta_{2}$ for
$l \gtrsim \rho_{\textrm{max}}$. Walks with $l(s) < \rho_{\textrm{max}}$ behave in a clearly different way, resulting in $\delta_{1} < \delta_{2}$.
This means that on the different length scales, different terms dominate in the summation of Eq.~\eqref{tausor} (see \refAppe{ap_dtauprop}).
A plausible argument is that as the $a$ exponent decreases in $P(\rho)$, the long-range links gain prevalence and the finite size of the system
affects its properties more. For instance, the mean value of link lengths is determined by $\rho_{\textrm{max}}$ in the $a \leq 2$ case
(the mean becomes infinite if $a \leq 2$ and $\rho_{\textrm{max}} \rightarrow \infty$). 

Based on this fact, one may argue that the $\delta_{2}$ exponent characterizes the system, and the $\delta_{1}$ exponent is not relevant. Still, 
the $l < \rho_{\textrm{max}}$ range can provide valuable information about the system, especially if we can identify which terms give relevant 
contribution to the behavior of $p_{0} (l)$ in Eq.~\eqref{tausor}.

\subsection{OpenStreetMap}

\begin{figure}
\begin{center}
\subfigure[Illustration of the OSM map of New York]{\label{fig_newyork}%
\includegraphics[width=\figurewidthA, angle=0]{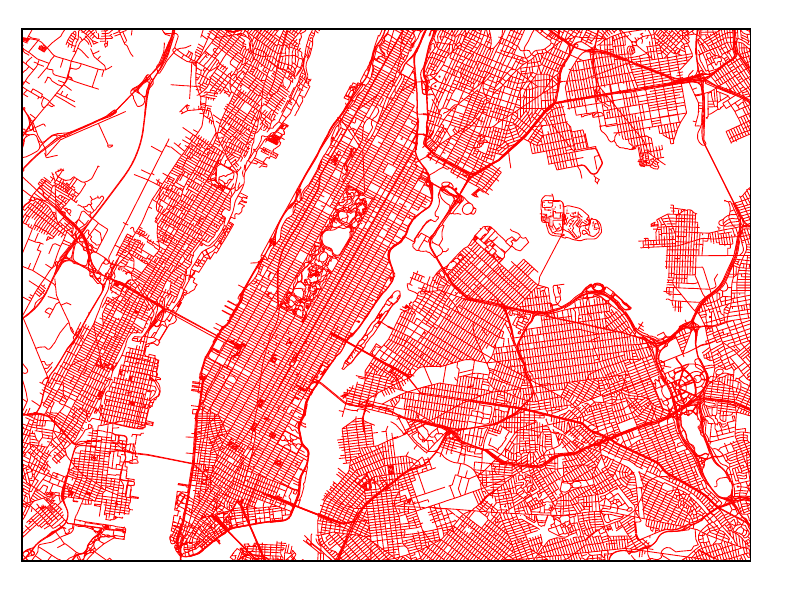}
}
\quad
\subfigure[Estimating $d_{\tau}$]{\label{fig_nytau}%
\includegraphics[width=\figurewidthA ]{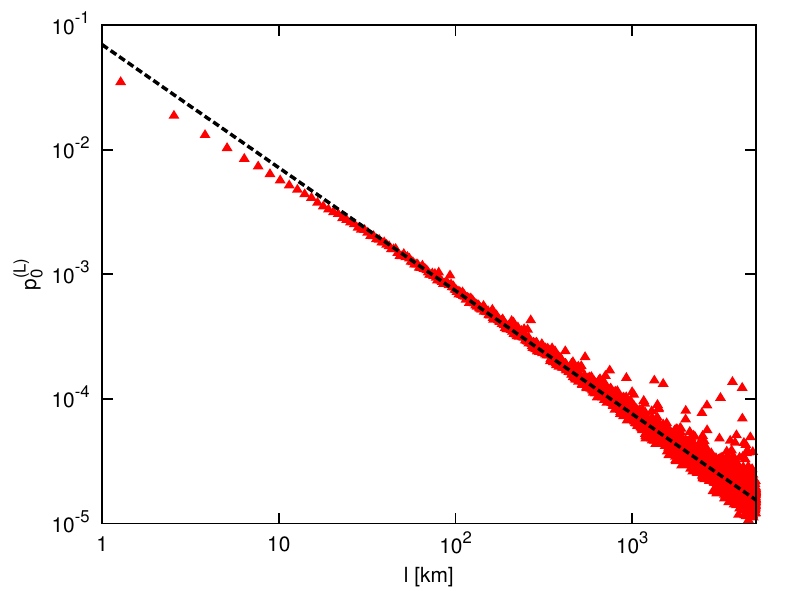}
}
\caption{Estimating the dimension of the New York City area road network. %
	(a) An excerpt of the network showing central New York. %
	(b) The distribution $P(l)$. The fit was obtained in the range
	$ l \in [ 7\,\mathrm{km}, 2981\,\mathrm{km} ] $. The estimated dimension is
	$d_{\tau} = 1.975 \pm 0.003 $ .}
\end{center}
\end{figure}

We used the road network of several cities and countries as a test case for the dimension measures presented previously.
We obtained the data from the OpenStreetMap (OSM) database, which is a large collaborative project to create a free editable map of the
world \cite{osm_web}. Due to the self-organizing nature of the project, OSM data are freely available in various flavors. We used several maps from
which we extracted the underlying road network 
\footnote{As raw OSM networks contain rich auxiliary information (e.g. the outline of buildings) and also introduce network nodes to describe the 
curvature of road segments, we applied a thorough data cleaning procedure to eliminate these artifacts and obtain the clean road network.}.
We regard the OSM networks as benchmarks of the dimensionality measurements, as locally they can be regarded as two dimensional lattices.
Hence, we expect these networks to have estimated dimensions close to $2$.

\reffig{fig_newyork} shows the road network of New York City's central urban area, obtained from the OpenStreetMap database. Note that the 
calculations were done on the entire New York area map, including suburbs and outskirts, which were omitted from this illustration. 
In \reffig{fig_nytau} we depict the obtained scaling for $d_{\tau}$ for the road network of the New York City area, an example where many parts of the 
network are close to a regular lattice. \refTab{tab_map} shows the obtained dimensions for the several road networks considered. Not surprisingly, we
have found that all three methods result in dimensions close to $2$.

To inspect the sensitivity of the dimension measures, we introduced a modified version of the road maps, where we removed all nodes with degree $2$
(i.e.~if $i$ is a node with $d_{i} = 2$ and with links $(i,j)$ and $(i,k)$ then we remove $i$ and add the link $(j,k)$ instead, recursively).
Typically, these 2-degree nodes are included for the proper visualization of a road (e.g.~a curved segment is actually stored as a sequence of chords),
but do not carry information about the topological structure of the actual network. Our processing procedure can be interpreted as a coarse-graining
step that reflects the road topology more precisely, while fading out the fine-resolution spatial details of the network. To see how this change in
resolution affects the properties of the random walk processes, we also give the resulting dimension values in \refTab{tab_map}. For the majority of
the maps there is no significant change in the estimated dimensions. The cleaning procedure only slightly modifies the exponents; in several cases,
the quality of fits is better however. It can also be observed that $d_{\rho}$ tends to produce higher values than $d_{s}$ and $d_{\tau}$, and that
it slightly overestimates the embedding dimension of the road networks.

\begin{table}
\centering
     \newcolumntype{R}{>{\centering\arraybackslash}X}%
     \begin{tabularx}{8cm}{l||R R R||R R R}
          \multirow{2}{*}{City / Region} & \multicolumn{3}{c||}{original} & \multicolumn{3}{c}{processed} \tabularnewline
	  & $d_{s}$ & $d_{\rho}$ & $d_{\tau}$  & $d_{s}$ & $d_{\rho}$ & $d_{\tau}$ \tabularnewline \hline \hline
Boston & 2.0 & 2.2 & 1.9 & 1.9 & 2.1 & 1.9 \tabularnewline
Budapest & 2.0 & 2.0 & 1.7 & 1.9 & 1.8 & 1.8 \tabularnewline
New York City & 2.1 & 2.3 & 2.0 & 2.0 & 2.2 & 1.9 \tabularnewline
Rome & 2.0 & 2.2 & 2.0 & 2.1 & 2.3 & 2.0 \tabularnewline
\hline \hline
Connecticut & 2.2 & 2.4 & 2.0 & 2.1 & 2.3 & 2.0 \tabularnewline
Estonia & 2.0 & 2.2 & 1.8 & 2.0 & 2.3 & 1.8 \tabularnewline 
New Mexico & 1.7 & 2.0 & 1.6 & 1.9 & 2.2 & 1.9 \tabularnewline
     \end{tabularx}
     \caption{The estimated dimensions (best fit) for several road networks. We depict the results for four metropolitan areas and three larger scale maps.
		Calculated statistical error of the exponents is approximately $0.05$.}
     \label{tab_map}
     
\end{table}

\subsection{Internet infrastructure}
\label{sec_tr}
We also study the Internet router network, which was 
obtained from a global topology discovery campaign presented in~\cite{matray}. The locations of the nodes were measured by the 
Spotter method~\cite{spotter} which gives a median error of $\approx 30 \, \mathrm{km}$ providing sufficient city level precision. To minimize the 
effect of mislocalizations a thorough consistency test was applied on the collected data. The data set contains 13,120 addresses which are 
considered well positioned. Between these, there are 44,116 links for which both endpoints have reliable location information.

\begin{figure}[tbp]
	\begin{center}
	\includegraphics[width=\figurewidthA, angle=0]{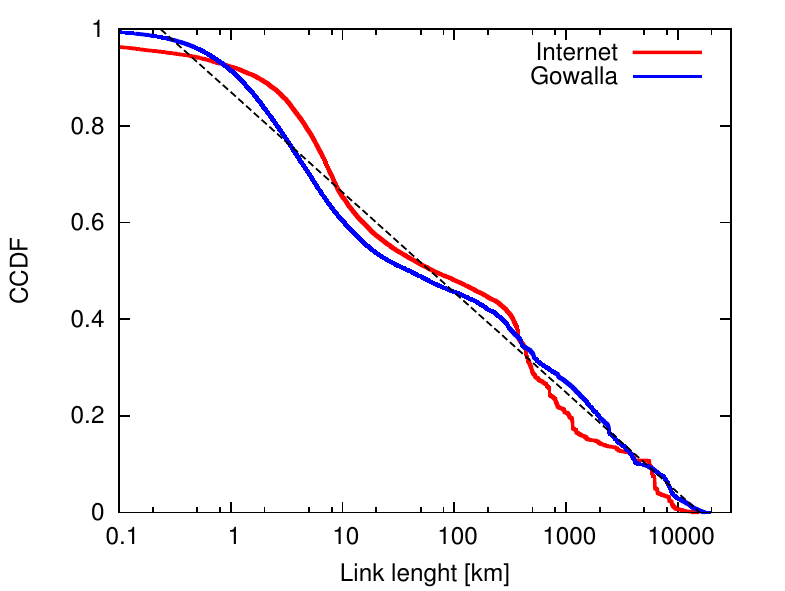}
	\caption{The distribution of link lengths for the Gowalla geosocial network and the Internet. In both cases the CCDF shows a linear decay over 4 
	orders of magnitude on the log-lin plot, indicating that $P(\rho) \sim 1 / \rho$. A more detailed analysis for the Internet dataset was
	already presented in~\cite{matray}. The Pearson correlation coefficients are $C_\mathrm{Gowalla} = 0.994$ and $C_\mathrm{Internet}=0.991$.}
	\label{fig_lld}
\end{center} 
\end{figure}

For our Internet data set, the distribution of link lengths can be approximated as $P(\rho) \sim 1 / \rho$ over four orders of magnitude
(see \reffig{fig_lld}) and also~\cite{matray}).

The results of the random walk processes are shown in \reffig{fig_gwtr} for the three dimension measures. The spectral dimension shows scaling,
although only for a rather narrow range. The resulting dimension is $d_s = 3.06$, which is higher than the embedding dimension.

The $d_{\rho}$ dimension is not applicable, as instead of the power-law relationship we get an exponential decay, as 
it is shown on the lin-log plot of \reffig{fig_gwtrdq}. Again, we can interpret this result as an infinite $d_{\rho}$ dimension.

Considering the $d_{\tau}$ dimension, similarly to the $a = 1$ case in the synthetic networks, we have two scaling regimes, which are again 
separated by the approximate size of the system (now the longest link can be at most $\approx 20,000 \, \mathrm{km}$ long).
We have $\delta_{1} \approx 1$, implying $d_{\tau} \approx 2$. While this
would be an appealing result as $d_{\tau}$ would reproduce the embedding dimension, we note, that this can also be readily explained by arguments
based on the link length distribution $P(\tau)$. Considering Eq.~\eqref{tausor} (see \refAppe{ap_dtauprop}), if we only take the first term
($p_{2}$) into account, we get $P(l) \sim P(\tau = l/2) \sim 1 / l$ in this case. This means, 
that the behavior observed here can be explained by the argument that all the other terms give only insignificant contribution on 
length scales $100\,\mathrm{km} \lesssim l \lesssim 20,000\,\mathrm{km}$, and thus the scaling is dominated by the term proportional to the 
distribution of link lengths.

For the larger length scales, we get $\delta_{2} = 1.44$, giving $d_{\tau} = 2.87$, which is larger than the embedding dimension.

\subsection{Gowalla geosocial network}
\label{sec_gow}

\begin{figure}
\begin{center}
\subfigure[Estimating $d_s$]{\label{fig_gwtrsp}%
\includegraphics[width=\figurewidthA]{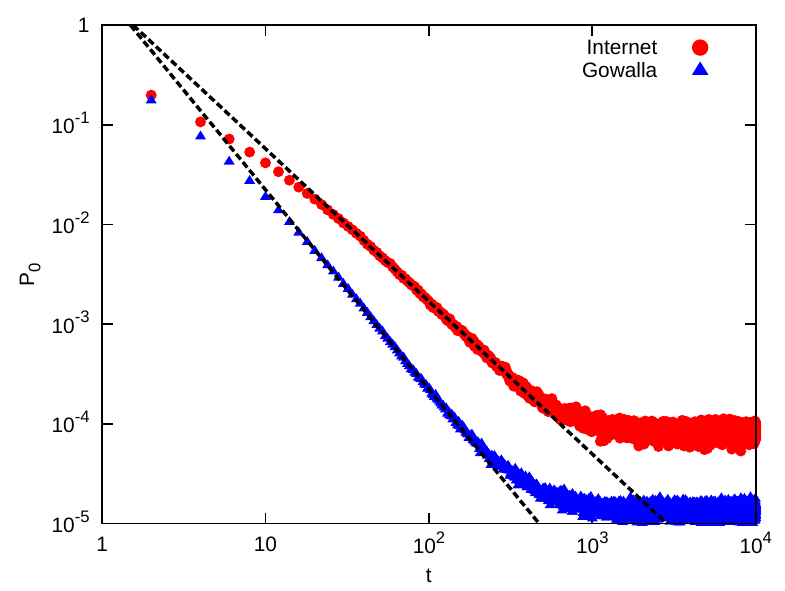}
}
\,
\subfigure[Estimating $\,d_{\rho}$]{\label{fig_gwtrdq}%
\includegraphics[width=\figurewidthA]{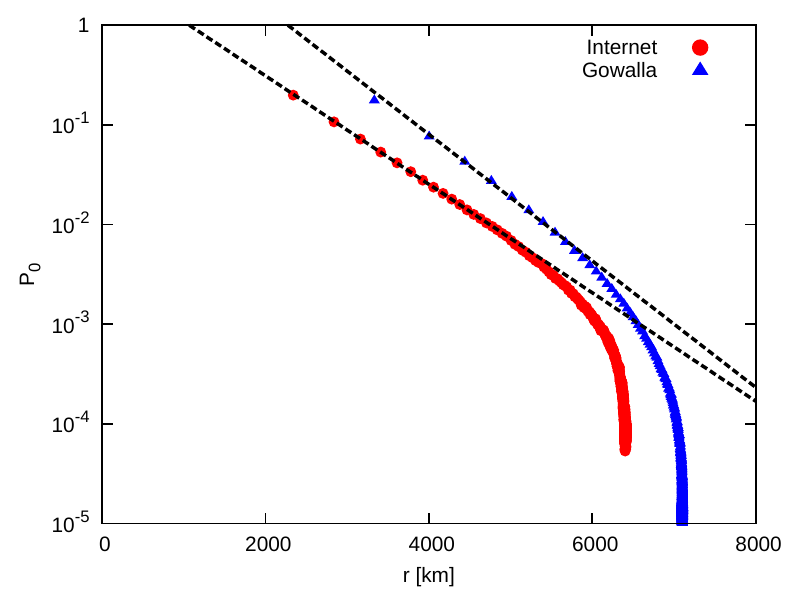}
}
\\
\subfigure[Estimating $d_{\tau}$ for the Internet infrastructure]{\label{fig_trtau}%
\includegraphics[width=\figurewidthA]{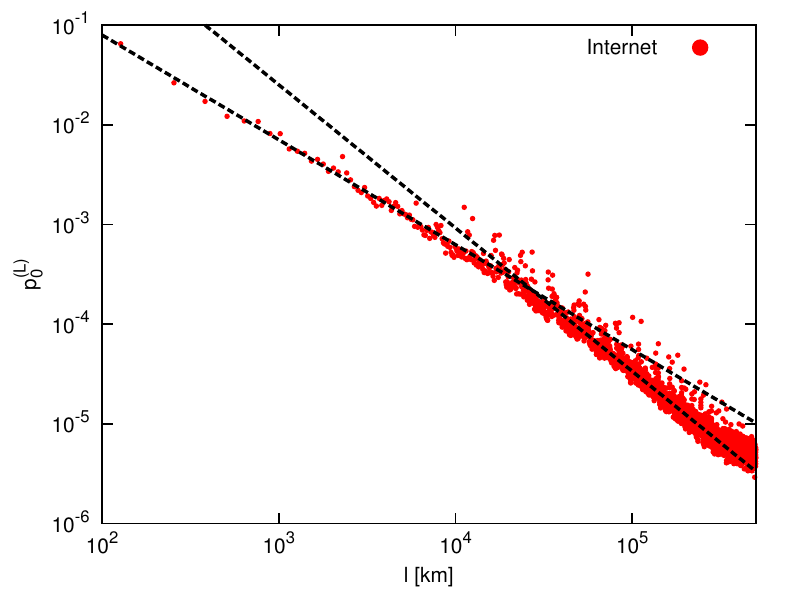}
}
\,
\subfigure[Estimating $d_{\tau}$ for the Gowalla geosocial network]{\label{fig_gwtau}%
\includegraphics[width=\figurewidthA]{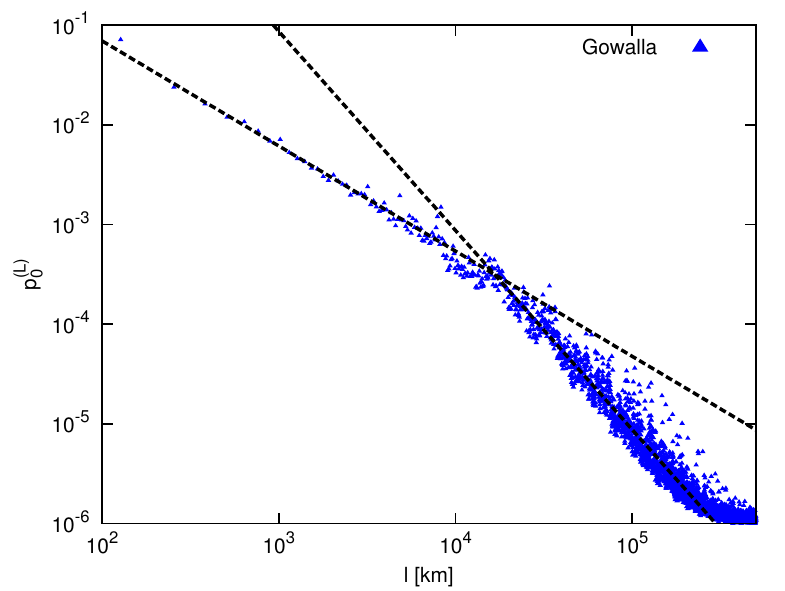}
}
\caption{Comparing the different dimension measures for the Internet and the Gowalla geosocial network. %
	(a) For $d_s$ the network exhibits a scaling behavior only in a limited regime with estimated
	spectral dimension $d_{s}^{(I)} = 3.06$ and $d_{s}^{(G)} = 4$. The fitting ranges are $t \in [29,330]$ and
	$t \in [15,148]$ for the Internet and Gowalla datasets respectively.%
	(b) The $d_{\rho}$ dimension measure for the Internet and the Gowalla geosocial network. %
	In this case we cannot identify a scaling behavior in any of the networks;
	the beginning of the curves seems to exhibit an exponential decay
	(note that the $x$ axis is linear and the $y$ is logarithmic). %
	(c) The $d_{\tau}$ dimension measure for the Internet infrastructure network. %
	Similarly to the synthetic case, we have two distinct scaling regimes, and the scaling behavior changes at approximately the
	size of the system. The dimensions are $d_{\tau,1}^{(I)} = 2.1 \pm 0.04$ for $l \lesssim \tau_{\textrm{max}}$ and
	$d_{\tau,2}^{(I)} = 2.87 \pm 0.02$ for $l \gtrsim \tau_{\textrm{max}}$. The fitting ranges are $l \in [148\,\mathrm{km},8103\,\mathrm{km}]$
	for $d_{\tau,1}^{(I)}$ and $l \in [36315\,\mathrm{km},268337\,\mathrm{km}]$ for $d_{\tau,2}^{(I)}$.
	(d) The $d_{\tau}$ dimension measure for the Gowalla geosocial network. Again, we have two scaling regimes.%
	In the first regime we get $d_{\tau,1}^{(G)} = 2.11 \pm 0.06$ for $l \lesssim \tau_{\textrm{max}}$, and we have
	$d_{\tau,2}^{(G)} = 3.99 \pm 0.02$ for $l \gtrsim \tau_{\textrm{max}}$. The fitting ranges are $l \in [544\,\mathrm{km},8103\,\mathrm{km}]$
	for $d_{\tau,1}^{(G)}$ and $l \in [22026\,\mathrm{km},268337\,\mathrm{km}]$ for $d_{\tau,2}^{(G)}$.
	}
\label{fig_gwtr}
\end{center}
\end{figure}

Gowalla was a geosocial network where users also provided spatial information via 
``check-ins'' (the service was shut down in 2012 \cite{gwfb}). We used the dataset described by Cho \emph{et al.}~\cite{gowalla}, which is freely 
available at the Stanford Large Network Dataset Collection website \cite{snap}. This data set includes 196,591 users (nodes) and 950,327 
(undirected) links (friendships), with a total of 6,442,890 check-ins, where a check-in is the position of a specific user at a specific time. The 
data set includes check-ins collected in the interval between Feb.~2009 and Oct.~2010.

As the location of these check-ins varies, we used a preprocessing technique to assign a unique position to the users (see \refappe{ap_gow}). In 
our analysis, we only included users, where the positions could be determined with sufficient confidence. This resulted a network 
which consists of 94,798 nodes with reliable position information and 289,961 links between them. Examining the distribution of link lengths, we 
get a behavior which is very similar to that observed for the Internet. Plotting the CCDF, it can be well approximated with $\ln(\rho)$, 
indicating that the PDF scales as $P(\rho) \sim 1 / \rho$  (\reffig{fig_lld}).

The evaluation of the previous dimension concepts is illustrated in \reffig{fig_gwtr} for the three dimension measures. The qualitative
behavior is quite similar to that obtained for the Internet, while there are quantitative differences. The spectral dimension is
$d_s = 4$, although the scaling regime is again rather limited. The $d_{\rho}$ dimension again does not produce scaling, but exhibits the
exponential decay also seen previously.

Considering the $d_{\tau}$ dimension, we can again identify the two scaling regimes, with the boundary at $\approx 20,000 \, \mathrm{km}$, the
approximate size of the system. Similarly to the Internet, we have $\delta_1 \approx 1$, giving $d_{\tau} \approx 2$ for the
$l \lesssim 20,000 \, \mathrm{km}$ range. For the longer walks, we get $\delta_2 = 1.997$ giving $d_{\tau} = 3.99$, which is again larger than
the embedding dimension, and is approximately the same as $d_s$.

Based on our results, we argue that some, but not all of the qualitative aspects of the real-world networks can be explained with the effect of 
the very broad $a = 1$ link length distribution. Similarly to the random networks, $d_{\tau}$ shows the same separation of length scales, and 
again we find that $d_{\rho}$ is not applicable here. However, for the investigated real-world networks, $d_{\rho}$ exhibits a clear exponential 
decay which was not the case for the synthetic network. Also, $d_{\tau}$ behaves in a quantitatively different way: the arising $\delta_2$ 
exponents differ significantly from the synthetic case, and also between the two real-world networks. Regarding the $d_{s}$ spectral dimension, we 
find significantly better scaling compared to the synthetic case. These observations mean that the link length distribution is not sufficient in 
itself to entirely control the dimensionality of complex networks, but further structural properties are also expected to have a traceable effect. 
In case of $d_{\tau}$, these hidden structural characteristics can be approached via the varying significance of the different terms in
Eq.~\refeq{tausor}. A rather interesting property is that for the $l \lesssim 20,000\,\mathrm{km}$ range both of our real-world networks give
$\delta_1 \approx 1$, implying $d_{\tau} \approx 2$, matching the embedding dimension of the system. In both networks, the link length distribution
can be well approximated by $P(\tau) \sim 1 / \tau$. Using this, the scaling $P(l) \sim 1 / l$ can be explained by the arguments presented in
\refSect{sec_tr}, based on Eq.~\refeq{tausor}. Still, the synthetic random networks produce different exponents, meaning that this peculiar
property arises only in both real-world networks of different origin.

\section{Discussion}
\label{sec_discussion}

In this paper we have introduced a generalization of the spectral dimension of networks.
Contrary to recently proposed methods to measure the dimension of spatially embedded
networks \cite{daqing}, our method does not utilize global information on the embedding
(i.e. the distance between any two nodes in the network). The key idea behind our method
is to exploit \emph{local}, link-wise information arising naturally from a random walk
process on the network. Consequently, the new dimension concept grasps the large-scale
structure as seen ``from the eyes of the random walker''. This means that the
real time dynamics of the random walk process is taken into account instead of only
using the abstract ``diffusion time'' (i.e. the number of steps taken). An appealing property of this
method is that it can be readily applied to \emph{partially embedded networks}, where
distances or delays are only available as a link-wise property and not generally. In the
case of a spatially embedded network, these delays can be a simple function of the distances
between the nodes, but we can utilize arbitrary delays possibly arising from physical
processes on the network. A consequence of this relaxation is that this link-wise property
does not need to qualify as a metric in a mathematical sense. Indeed, in many real-world
settings delays can violate the triangle inequality, as taking a ``detour'' may result in
less accumulated delay.

To compare our dimension concept ($d_{\tau}$) with the classical spectral dimension ($d_s$) and the method of Daqing \emph{et al.}~\cite{daqing}
($d_{\rho}$), we considered a spatially embedded random network, and different real-world networks: road networks, Internet routers and the Gowalla 
geosocial network. We used the road networks as benchmarks, since they substantially resemble a two dimensional lattice. As expected, all three 
methods provide dimensions $\approx 2$. We also studied the effect of coarse graining the road networks and found that in most cases the methods 
yield similar results. Our random networks were generated to have a scale-free distribution of link lengths: $P(\rho) \sim \rho^{-a}$. We found 
that for exponents $a \geq 2$ all methods exhibit scaling, and in accordance with \cite{daqing}, the value of $a$ directly affects the measured 
dimension of the network.

Contrary to previous works that we are aware of, in this paper we have also considered the theoretically and empirically designated case of $a=1$.
For this very broad distribution of link lengths, $d_s$ and $d_{\rho}$ decay faster than a power law, and thus can be considered infinite. 
Interestingly, $d_{\tau}$ remains finite, but there appears a separation of length scales with two different exponents. In accordance with several 
empirical studies \cite{yook,lambiotte,nowell,matray,distance,gowalla}, our Internet and Gowalla dataset can also be well characterized by $a 
\approx 1$. An important empirical result of our paper is that while for these networks $d_\rho$ can be considered infinite, $d_{\tau}$ remains
finite, with similar scaling regimes which are again separated by the approximate size of the system.

For the Internet, this seems to contradict the previous result of~\cite{daqing}, where $d_{\rho}^{(I)}$ was found to be $4.5$. A possible 
explanation for this mismatch lies in the nature of the data sets applied. For instance, there is a significant difference in the distribution of 
link lengths: in~\cite{daqing}, the authors find $a \approx 1.6$ \footnote{The $\delta$ exponent used in~\cite{daqing} corresponds to the 
conditional probability, what they estimate based on the assumption that the network nodes are uniformly distributed.}. We believe that due to the 
applied geolocalization technique, our data set is more accurate in terms of the geographic position of the nodes~\cite{spotter,matray}.

Comparing the results for the real-world and synthetic networks in the $a = 1$ case, we find a qualitatively similar behavior. While $d_{\rho}$ is 
not applicable, for $d_\tau$ there emerge two scaling regimes with different exponents, separated by the system size. On the other hand, the 
behavior of $d_s$ is apparently different: in real-world networks there appears a limited scaling regime, which is absent in the synthetic 
setting. Furthermore, even the behavior of $d_\tau$ is quantitatively different: the arising exponents differ both between the synthetic and real 
world case, and between the two real-world networks. This result implies that an uncorrelated random network model is not able to reproduce the 
structure of a real network with a similar $P(\rho)$. Thus, in spite of its distinguished significance, the link length distribution is not 
sufficient in itself to characterize the dimensionality of a complex network.

Our results show that the time dynamics of the diffusion is indeed a very important aspect. In the classical
case of diffusion, each step takes the same amount of time. In this case we can take the continuous time limit,
which gives a Markovian master equation for the evolution of the probability distribution~\cite{brspectrum}.
The spectral dimension in this case can also be extracted from the spectral density of the transition
matrix~\cite{spectral}. In the case of diffusion taking place on a network with non-uniform transmission delays
on the links, the continuous time limit would result in a delay-differential equation~\cite{delay}.
Further research could clarify the relation between the spectrum of arising from these equations and the
dimension concept introduced in our paper.

%
%
%
%

\section*{Acknowledgments}

The authors thank the partial support of the EU FP7 OpenLab project (Grant No.287581), 
the OTKA 7779 and 103244, and the NAP 2005/KCKHA005 grants.
This work was partially supported by the European Union and the European Social Fund through project FuturICT.hu
(grant no.: TAMOP-4.2.2.C-11/1/KONV-2012-0013).
The authors I.Cs. and G.V. thank the financial support of the MAKOG Foundation.


%
%
%
%

\appendix

\section{The connection between $P_0 (t)$ and $P(l)$}
\label{ap_p0}

We note, that the length distribution of returning walks ($P(l)$ in Eq.~\eqref{taudim}), and the arising dimension measure $d_{\tau}$, can be
considered the generalization of the $P_0 (t)$ return probability in Eq.~\eqref{spectraldim} and the $d_{s}$ dimension.

To show that, we now give a possible alternative definition of Eq.~\refeq{spectraldim} for finite networks, which can be generalized to give the 
scaling relation Eq.~\refeq{taudim} in a natural way. Let us define some upper limit $T$ and consider walks $s$ which have at maximum $T$ steps
($|s| \leq T$). Of course, for a finite network we will have a finite number of such walks. Now, among these walks, we shall consider those, that 
return to their origin:
\begin{equation}
	S_{0}^{(T)} \equiv \{ s: m \equiv |s| \leq T \, \mathrm{and} \, s_{0} = s_{m} \} \textrm{.}
\end{equation}
Among these, we can define the discrete probability distribution $p_{0}^{(T)} (t)$ as the relative abundance of walks of length $t$ in $S_{0}^{(T)}$.
For a finite network and a fixed $T$, the $p_{0}^{(T)} (t)$ distribution is simply the rescaled version of $P_{0} (t)$ in Eq.~\refeq{spectraldim}:
$p_{0}^{(T)}(t) = C^{(T)} P_{0} (t)$, where $C^{(T)}$ is an appropriate normalizing constant so that $\sum_{t=1}^{T} p_{0}^{(T)}(t) = 1$.
Hence, we obtain a scaling relation with the very same exponent as in Eq.~\refeq{spectraldim}:
\begin{equation}
	p_{0}^{(T)}(t) \sim t^{-\alpha} ,
	\label{p0t}
\end{equation}

This relation can be easily generalized to give an estimate of the scaling relation Eq.~\refeq{taudim}. In this case, instead of
limiting the number of steps taken, we will introduce a limit $L$ for the distances along the walk: we will consider walks $s$ which
have $l(s) \leq L$. Among these, we will again define the set of those walks which return to the origin, and consider their probability
distribution denoted by $p_{0}^{(L)} (l)$. As the choice of $L$ does not alter the obtained scaling exponents, we choose an $L$ value
appropriate for the system considered, and omit the superscript and write $P(l) \equiv p_{0}^{(L)} (l)$.

Note that now $l$ is a continuous variable, meaning that for a finite network, we have
\begin{equation}
	P(l) \sim \sum_{\substack{\textrm{walks}\,s \\ \textrm{where}\,s_{0} = s_{m}}} \delta \left (l - l(s) \right ) \textrm{,}
	\label{aml}
\end{equation}
where $\delta(x)$ is the Dirac delta function. Choosing an appropriate binning size, $P(l)$ and the scaling exponents can be well estimated numerically.

%
%
%
%

\section{Some properties of $d_{\tau}$}
\label{ap_dtauprop}
In accordance with \refAppe{ap_p0}, for some finite $L$ value we define $P(l)$ as the distribution of lengths $l(s)$ of returning walks $s$ with
$l(s) < L$. This distribution can be calculated by summing the individual distributions which arise from walks with a given number of steps: 
\begin{equation}
\label{tausor}	
	P(l) = \sum_{i=1}^{\infty} p_{i} (l)
\end{equation}
where $p_{i} (l)$ is the term corresponding to walks with $|s| = i$ number of steps and maximum cumulated length $L$. To give an example, the
first four terms in the sum are:

\begin{align}
	p_{1} (l) = & 0 \\[0.5em]
	p_{2} (l) = & \vtop{\vskip-3.5ex\hbox{\includegraphics[width=2em, angle=0]{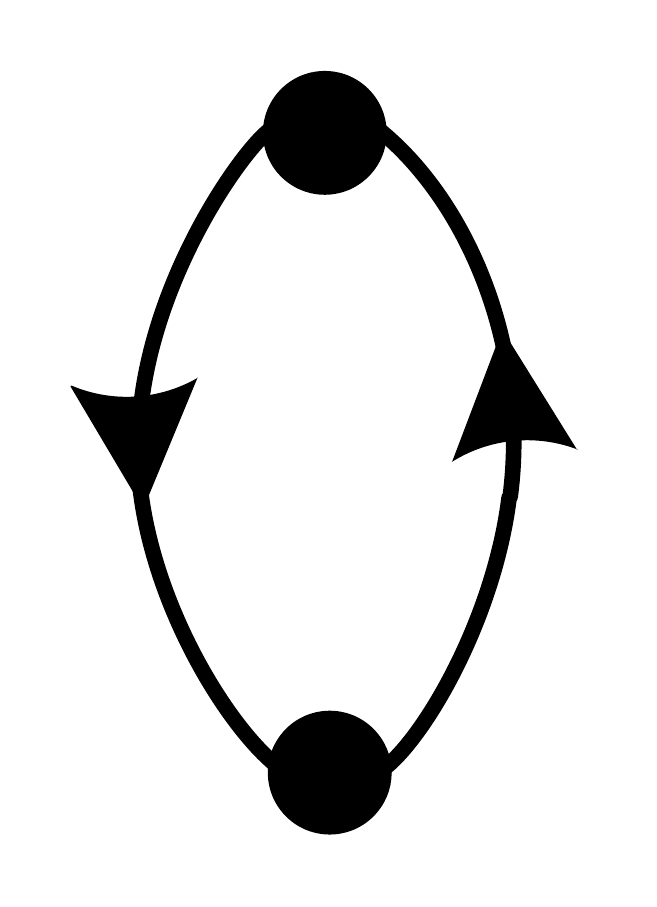}}} = \frac{1}{k} P(2 \tau = l) \label{tausor1} \\[0.5em]
	p_{3} (l) = & \vtop{\vskip-3.5ex\hbox{\includegraphics[width=2.5em, angle=0]{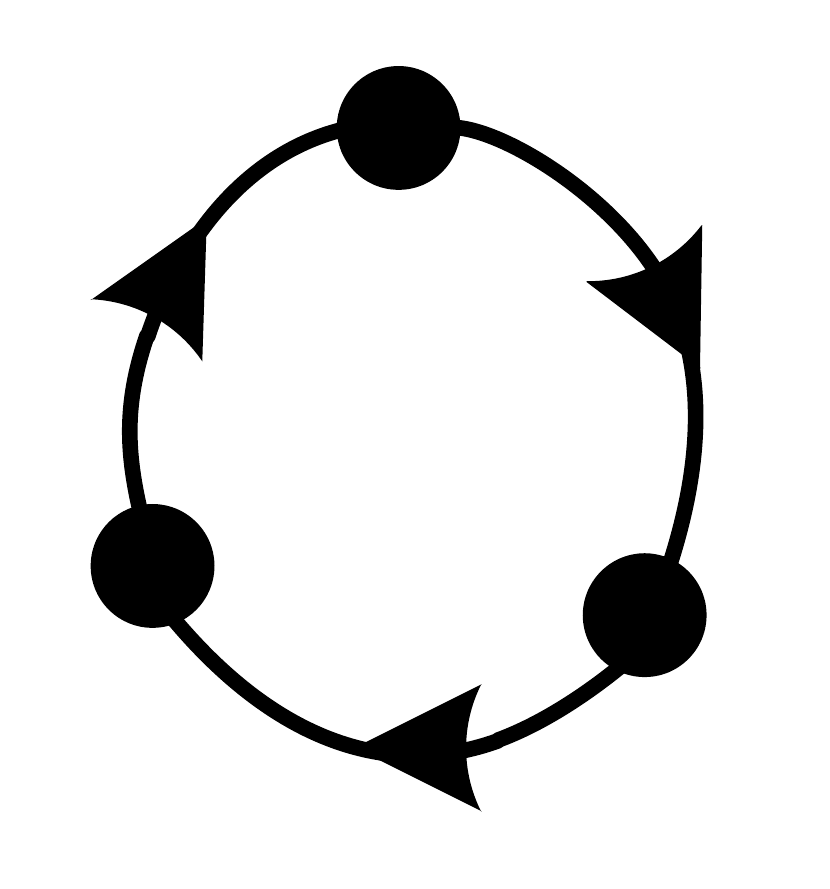}}} = \frac{C_{(3)}}{k^{2}} P(\tau_{1} + \tau_{2} + \tau_{3} = l) \label{tausor2} \\[1em]
		p_{4} (l) = & %
     \vtop{\vskip-3.7ex\hbox{\includegraphics[width=2em, angle=0]{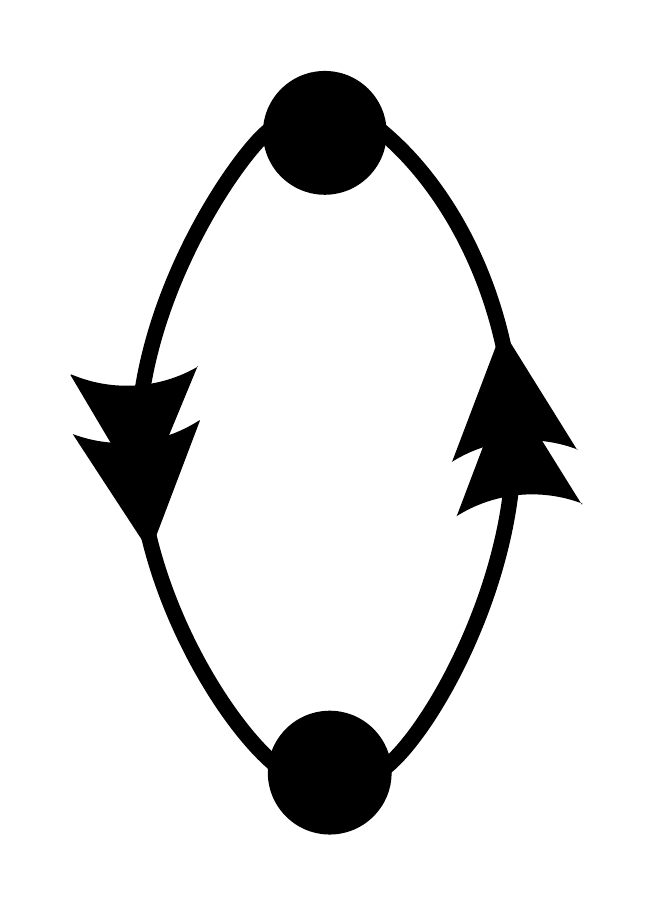}}} + %
     \vtop{\vskip-5ex\hbox{\includegraphics[width=1.55em, angle=0]{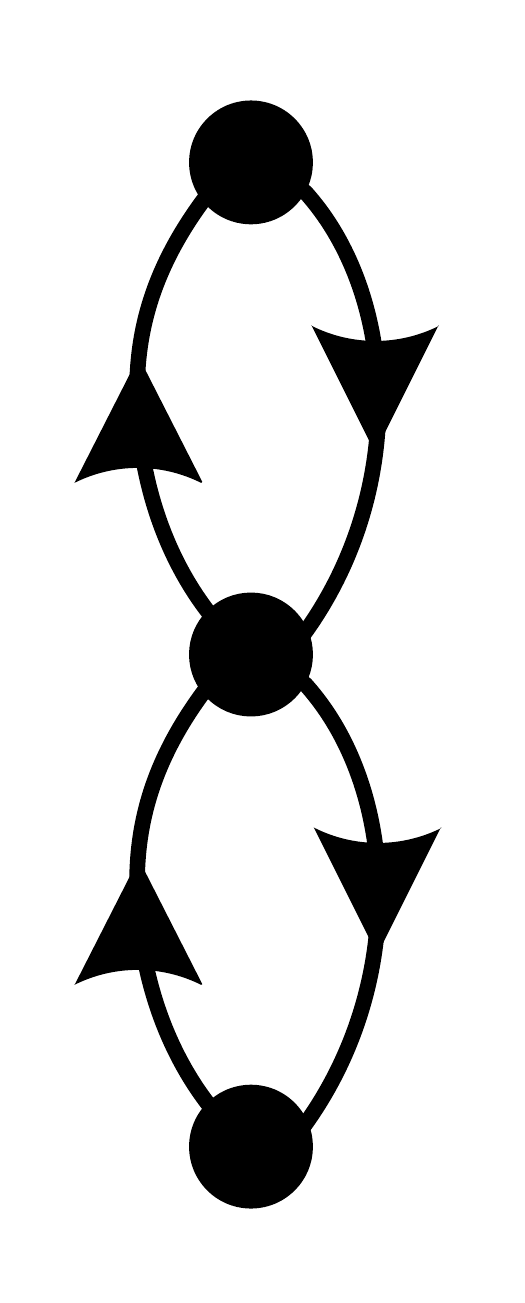}}} + %
     \vtop{\vskip-4ex\hbox{\includegraphics[width=2.75em, angle=0]{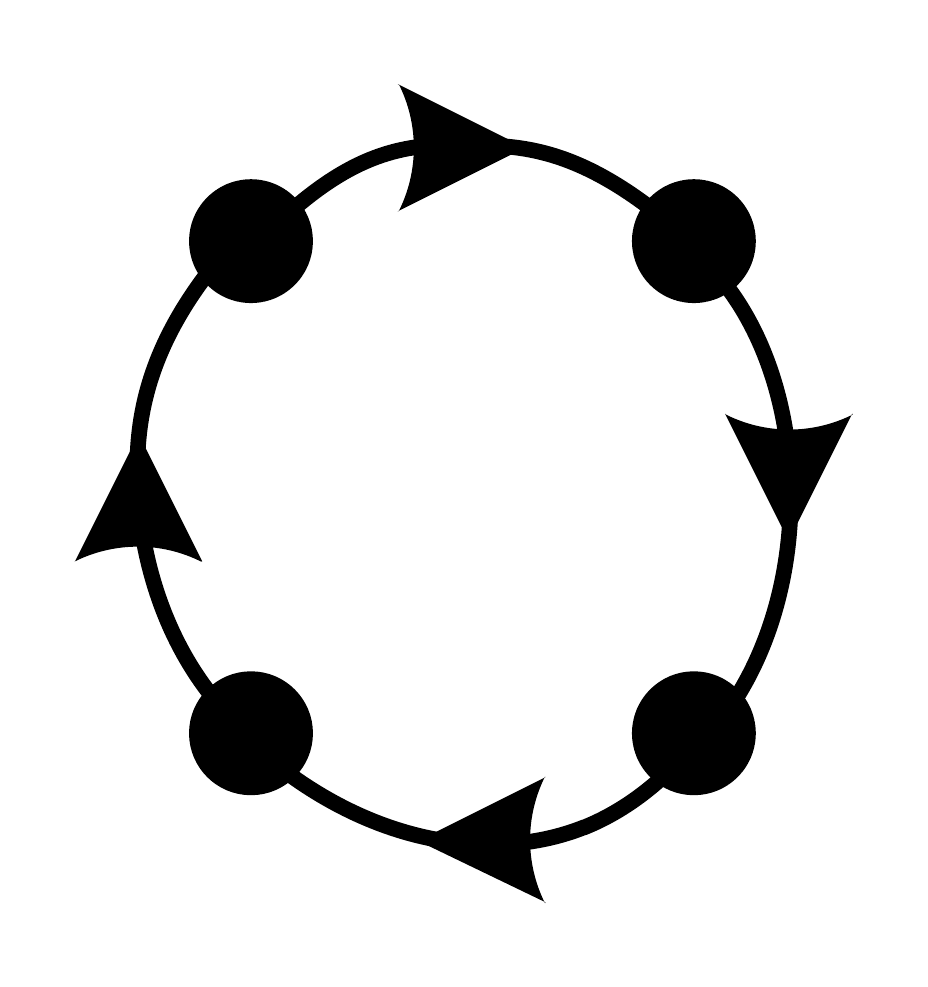}}} = %
     \nonumber \\ %
     = & \frac{1}{k^{3}} P(4 \tau = l) + 2 \frac{k-1}{k^{3}} P(2(\tau_{1} + \tau_{2}) = l) + %
     \nonumber \\ %
     &	\frac{C_{(4)}}{k^{3}} P( \tau_{1} + \tau_{2} + \tau_{3} + \tau_{4} = l ) \label{tausor3}
\end{align}

Here, $k$ denotes the average node degree, $C_{(i)}$ is the probability of finding a circle of $i$ nodes (e.g.~$C_{(3)}$ is the clustering 
coefficient) and $P(\tau)$ denotes the distribution of link lengths (see also \refSect{sec_lld}). The higher order terms (e.g.~$P(2(\tau_{1} + 
\tau_{2}) = l)$ in Eq.~\eqref{tausor3}) correspond to the distribution of the sum of the lengths of two or more links, which can be calculated by 
the convolution of $P(\tau)$. This is only true in an uncorrelated network. In a real-world setting, the distribution of link lengths along 
circles of a specific length may from the overall link length distribution. For instance, in case of $C_{(3)}$, geographically 
concentrated triangles are expected to be much more prevalent than ones spanning large distances. In this case, the terms in Eq.~\refeq{tausor} 
can only be determined by the numerical analysis of the specific network at hand. Still, we can make some general statements about the relevance 
of the different terms.

In a network, where all link lengths are positive, we expect that the consecutive terms in Eq.~\eqref{tausor} will be centered around increasing 
$l_{i}$ values (with $l_{i} \sim i l_{1}$).
Also, there is certainly a minimum and a maximum length that can be spanned in $i$ steps: $p_{i} (l) \equiv 0$ if $l < i \tau_{\textrm{min}}$ or 
$l > i \tau_{\textrm{max}}$, where $\tau_{\textrm{min}}$ and $\tau_{\textrm{max}}$ denote the minimum and maximum link length in the network, 
respectively. This means that for a finite $L$, we will only have a finite number of nonzero terms in Eq.~\eqref{tausor}. More interestingly, for 
each $\lambda < L$ length, there will be an $i_{0}$ value, that the terms after $i_{0}$ only give decreasing contribution to $p_{0} (\lambda)$. 
This implies that for some $i_{1} \geq i_{0}$, we can neglect the terms with $i > i_{1}$. Thus, for a small $\lambda$ value, we have to take into 
account only a few terms in Eq.~\eqref{tausor}, and the behavior of $p_{0} (l)$ in the $l < \lambda$ case can be well approximated by examining 
the behavior of these few terms. In the case of the real-world networks considered in Sects.~\ref{sec_tr} and~\ref{sec_gow}, the behavior of
$P(l)$ for small $l$ values can be readily explained by the $P(\tau)$ link length distribution alone.

%
%
%
%

\section{Data processing for the Gowalla network}
\label{ap_gow}

The original Gowalla data set \cite{snap} consists of a social network (where nodes represent users and links symbolize friendships between them) 
and check-in data with approximately $6.5$ million records of the form: \texttt{(userID, timestamp, lat, lon)}. As most of the users performed 
check-ins from various locations, we need to designate a single base-location for each user, to enable the spatial embedding of the social 
network. We have chosen to reject the spatial embedding of users who does not seem to have a characteristic position over the check-ins.

To perform the embedding, we ``pixelate'' check-in locations of a given user with a hierarchical spherical indexing technique, the Hierarchical 
Triangular Mesh (HTM) \cite{htm}. At a given resolution level HTM divides the sphere into triangles, referred to as \emph{trixels}. We use base 
indices at depth 20, which corresponds to trixels of area $\approx 100 \textrm{m}^2$ or $\approx 1000 \textrm{ft}^2$. Next, for each user $u$ we 
iteratively coarse-grain the trixel-resolution (by decreasing the index depth) until there emerges a single trixel containing at least $50\%$ of 
$u$'s check-ins. We calculate the $\mu_u$ mean position and $\sigma_u$ standard deviation within this cell, and accept $\mu_u$ as the position 
estimate for $u$ if $\sigma_u < 10$ km, otherwise we reject the spatial embedding of $u$. After processing the network with this procedure there 
remains $94,798$ nodes and $289,961$ links, for which both endpoints have a well-defined location estimates.

%
%
%
%



\begin{thebibliography}{999}

\bibitem{rgcrit} M. E. Fisher,
				The renormalization group in the theory of critical behavior.
				{\em Rev. Mod. Phys.} {\bf 46} (4) {\em 597--616} (1974)

\bibitem{dyncrit} P. C. Hohenberg and B. I. Halperin,
	Theory of dynamic critical phenomena.
	{\em Rev. Mod. Phys.} {\bf 49} (3) {\em 435--479} (1977)
				
\bibitem{forest} B. Drossel and F. Schwabl
				Self-organized critical forest-fire model.
				{\em Phys. Rev. Lett. } {\bf 69} {\em 1629--1632} (1992).
				
\bibitem{traffic} A. J. da Silva and B. Sto\v sic
				Critical density of urban traffic.
				arXiv:1009.2180

\bibitem{critnet} S. N. Dorogovtsev, A. V. Goltsev and J. F. F. Mendes
				Critical phenomena in complex networks.
				{\em Rev. Mod. Phys.} {\bf 80} (4) {\em 1275--1335} (2008)

\bibitem{qg} J.~Ambj{\o}rn, J.~Jurkiewicz, and R.~Loll.
The Spectral Dimension of the Universe is Scale Dependent.
{\em Phys. Rev. Lett.} {\bf 95} (17) {\em 1--4} 
(2005)

\bibitem{spectral} S. Alexander, and R. Orbach,
				Hausdorff and spectral dimension of infinite random graphs.
				{\em J. Phys. (Paris) Lett.} {\bf 43} {\em 625} (1982).

\bibitem{hwang} Hwang, S., Yun, C.-K., Lee, D.-S., Kahng, B., and Kim, D. 
		Spectral dimensions of hierarchical scale-free networks with weighted shortcuts.
				{\em Phys. Rev. E} {\bf 82} {\em 1–12} (2010).
	
\bibitem{spectral2} S. Bilke, and C. Peterson 
	Topological properties of citation and metabolic networks.
				{\em Phys. Rev. E.} {\bf 64} {\em 036106} (2001). 

\bibitem{fractality} C. Song. S. Havlin, and H. A. Makse
				Origins of fractality in the growth of complex networks.
				{\em Nat. Phys.} {\bf 2} {\em 275} (2006).
	
\bibitem{graph_boxcount} C. Song, L. K. Gallos, S. Havlin, and H. A. Makse,
				How to calculate the fractal dimension of a complex network: the box covering algorithm.
				{\em J. Stat. Mech. Theory and Experiment } {\bf 03} {\em 03006} (2007).
		
\bibitem{fractal_sw} F. Kawasaki, and K. Yakubo
				Reciprocal relation between the fractal and the small-world properties of complex networks.
				{\em Phys. Rev. E } {\bf 82} {\em 036113} (2010).

\bibitem{daqing} L. Daqing, K. Kosmidis, A. Bunde, and S. Havlin
				Dimension of spatially embedded networks.
				{\em Nat. Phys.} {\bf 7} {\em 481-484} (2011).

\bibitem{diffnet} I.~Simonsen
				Diffusion and networks: A powerful combination!
				{\em Phys. A} {\bf 357} (2) {\em 317-330} (2005)

\bibitem{spotter} S. Laki, P. M\'atray, P. H\'aga, T. Sebok, I. Csabai, G. Vattay
				Spotter: a model based active geolocation service.
				{\em Proc. IEEE INFOCOM} (2011).

\bibitem{chaosbook} P. Cvitanovi\'c, R. Artuso, R. Mainieri, G. Tanner and G. Vattay,
				Chaos: Classical and Quantum.
				{\em ChaosBook.org, Niels Bohr Institute} (2009).

\bibitem{barthelemy} M. Barth\'el\'emy,
				Spatial networks.
				{\em Physics Reports} {\bf 499} {\em 1-101} (2011).

\bibitem{yook} S. H. Yook, H. W. Jeong, A. L. Barabasi,	
				Modeling the  Internet's  large-scale  topology.
				{\em Proc. Natl. Acad. Sci.} {\bf 99} {\em 13382-86} (2001).
		
\bibitem{lambiotte} R. Lambiotte et al.,
				Geographical dispersal of mobile communication networks.
				{\em Phys. A} {\bf 387} {\em 5317-5325} (2008).
\bibitem{nowell} D. Liben-Nowell, J. Novak, R. Kumar, P. Raghavan, and A. Tomkins,
				Geographic routing in social networks.
				{\em Proc. Natl. Acad. Sci.} {\bf 102} {\em 11623-11628} (2005).
		
\bibitem{gowalla} E. Cho, S. A. Myers, J. Leskovec,
				Friendship and Mobility: Friendship and Mobility: User Movement in Location-Based Social Networks.
				{\em Proc. ACM SIGKDD} (2011). 

\bibitem{matray} P. M\'atray, S. Laki, P. H\'aga, G. Vattay, and I. Csabai,
				On the spatial properties of internet routes.
				{\em Comp. Netw.} {\bf 56} {\em 2237-2248} (2012).


\bibitem{distance} J. Goldenberg, M. Levy,
		Distance is not dead: Social interaction and geographical distance in the internet era.
				arXiv:0906.3202
	
\bibitem{correldim} P. Grassberger, and I. Procaccia,
				Characterization of strange attractors.
				{\em Phys. Rev. Lett.} {\bf 50} {\em 346} (1983).

\bibitem{havlin_diffusion} S. Havlin, and D. Ben-Avraham,
				Diffusion in disordered media.
				{\em Adv. Phys.} {\bf 36 (6)} {\em 695-798} (1987).
\bibitem{brspectrum} A. Bray, G. Rodgers, 
				Diffusion in a sparsely connected space: A model for glassy relaxation.
				{\em Phys. Rev. B} {\bf 38 (16)} {\em 461–470} (1988).

\bibitem{derenyi} I. Farkas, I. Der\'enyi, A.-L. Barab\'asi, and T. Vicsek,
Spectra of “real-world” graphs: Beyond the semicircle law.
{\em Phys. Rev. E.} {\bf 64} (2) {\em 1-12} (2001).
 
\bibitem{spectra2}
Z.~Zhang, Y.~Qi, S.~Zhou, Y.~Lin, and J.~Guan,
Recursive solutions for Laplacian spectra and eigenvectors of a class of growing treelike networks.
{\em Phys. Rev. E.} {\bf 80} (1) {\em 1-6} (2009).

\bibitem{scalefreespectra}
D.~Kim and B.~Kahng,
Spectral densities of scale-free networks.
{\em Chaos} {\bf 17} (2) {\em 026115} (2007).

\bibitem{kleinberg} J. M. Kleinberg,
				Navigation in a small world.
				{\em Nature} {\bf 406} {\em 845} (2000).


\bibitem{humodell} Y. Hu, Y. Wang, D. Li, S. Havlin, Z. Di
    Possible Origin of Efficient Navigation in Small Worlds.
				{\em Phys. Rev. Lett.} {\bf 106} {\em 108701} (2011).


\bibitem{kosmidis} K. Kosmidis, S. Havlin, and A. Bunde,
				Structural properties of spatially embedded networks.
				{\em Europhys. Lett.} {\bf 82} {\em 48005} (2008).


\bibitem{osm_web} OpenStreetMap,
				\texttt{http://www.openstreetmap.org}

\bibitem{gwfb} 
Original website: \texttt{http://www.gowalla.com}; as of August 2012, the \texttt{gowalla.com} domain is
no longer available, see for example (the version cached by Google): \\
\texttt{http://webcache.googleusercontent.com/search?q=cache:blog.gowalla.com/post/13782997303/gowalla-going-to-facebook}

\bibitem{snap} Stanford Large Network Dataset Collection,\\
				\texttt{http://snap.stanford.edu/data}


\bibitem{htm} A. Szalay, J. Gray, Gy. Fekete, P. Kunszt, P. Kukol, and A. Thakar,
				Indexing the Sphere with the Hierarchical Triangular Mesh.
				{\em Microsoft Research Technical Report, MSR-TR-2005-123} (2005).			
				

\bibitem{delay}
P.~Weng,
Asymptotic speed of propagation of wave fronts in a lattice delay differential equation with global interaction.
{\em IMA J Appl Math} {\bf 68} (4) {\em 409--439} 
(2003).

\end{thebibliography}
\end{document}